%% file: n3960_flames.tex
\documentclass{aa}  
\usepackage{graphicx}
\usepackage{txfonts}
\usepackage{longtable}
\usepackage{slashbox}
\usepackage{lscape}
\usepackage{longtable,lscape}
\usepackage{amssymb}
\usepackage{natbib}
\bibpunct{(}{)}{;}{a}{,}{,}
\usepackage{times}

\def\Min{${}^{\prime}$\llap{.}}
\def\Sec{${}^{\prime\prime}$\llap{.}}
\def\deg{${}^\circ$}
\def\min{${}^{\prime}$}

\newcommand{\lithiumline}{Li\,{\footnotesize I} 6707.8\,$\AA$}

\begin{document}
   \title{Lithium abundances in the old open cluster NGC\,3960
   from VLT/FLAMES observations
\thanks{Based on observations collected at ESO-VLT, Paranal Observatory,
Chile, Programme numbers 73.D-0520(A)}}

   \subtitle{}

   \author{L. Prisinzano
          \inst{1}
          \and
          S. Randich\inst{2}
          }

   \offprints{L. Prisinzano}

   \institute{INAF - Osservatorio Astronomico di Palermo, 
   Piazza del Parlamento 1, I-90134 Palermo, Italy\\
              \email{loredana@astropa.inaf.it}
         \and
             INAF - Osservatorio Astronomico di Arcetri, 
	     Largo E. Fermi 5, I-50125 Firenze, Italy\\
             \email{randich@arcetri.astro.it}
             }

   \date{Received July 24, 2007; accepted September 7, 2007}

 
  \abstract
   {Old open clusters are very useful targets to investigate mechanisms 
   responsible for 
   lithium (Li) depletion during the main sequence. Comparison of the Li
   abundances in clusters of different age allows us to understand 
   the efficiency
   of the Li destruction process.}
   {Our goal is the determination of membership and Li abundance in
   a sample of candidate members of the open cluster
   NGC\,3960 (age $\sim$1\,Gyr), with the aim to fill the gap between
   0.6 and 2 \,Gyr in the  empirical description
of the behavior of the average  
 Li abundance as a function of the stellar age.
    }
   {We use VLT/FLAMES Giraffe spectra to determine the radial velocities and
   thus the membership of a
   sample of 113 photometrically selected 
   candidate cluster members. From the analysis of the Li line 
   we derive Li abundances for both cluster members and non-members.}
   {  stars have radial velocity consistent with membership, 
  with an expected  fraction of contaminating field
   stars of about 20\%. Li is detected in
   29 of the RV members; 
   we consider these stars as cluster members, while
 we make the reasonable assumption that the remaining 10 RV members without Li, 
 are among the contaminating stars. 
    Li abundances of the stars hotter than about 6000\,K
   are similar to those of stars in the Hyades, while they are slightly
    smaller for cooler stars. This confirms that NGC\,3960 is
     older than the Hyades.}
   {The average Li abundance of stars cooler than about 6000\,K indicates
   that the Li Pop. I plateau   might  start already at $\sim$1\,Gyr 
   rather than 2\,Gyr that
   is the upper limit previously derived in the literature. We also find that
   the fraction of field stars with high Li abundance ($\gtrsim $1.5) is about
   one third of the whole sample, which is in agreement with previous estimates.
   The fraction of contaminating field stars is consistent with
   that previously derived by us from photometry.}

\titlerunning{Lithium abundances in the old open cluster NGC\,3960}
   \maketitle
%

\section{Introduction\label{introdu}}
Open cluster (OCs) are commonly recognized as one of the best tools
to investigate the formation and evolution of the Galactic disk,
as well as the evolution of stars and their properties.

We have carried out a VLT/FLAMES project aimed at deriving homogeneous
information for a sample of 11 old OCs (ages greater than about
0.9~Gyr) with the aim of a) deriving their chemical composition
from UVES spectra of evolved stars; b) determine radial velocities
and thus membership, together with lithium (Li) abundances from Giraffe spectra
of turn-off (TO) and main sequence (MS) cluster candidates 
\citep{rand05,pall06}.
In this paper we focus on Giraffe observations of
NGC\,3960, the youngest OC in our sample. 

In the last few years this cluster has been object of
different photometric and spectroscopic studies, that have allowed
a more accurate determination of its parameters. \citet{brag06}
 derived an age between 0.6 and 0.9\,Gyr, a distance modulus
(m-M)$_0=11.6 \pm 0.1$, a reddening E(B--V)=0.29$\pm 0.02$, with differential
reddening $\Delta$E(B--V)=0.05. A slightly older 
age (in the range 0.9--1.4\,Gyr) was derived by \citet{pris04} who
also estimated the cluster mass function from the luminosity
function in the V and J bands,
finding a slope $\alpha=2.95 \pm 0.53$ for 
masses above 1~M$_{\odot}$. Note that the contamination from field stars
was taken into account using a control field and performing a statistical
subtraction. A spectroscopic [Fe/H]=$-0.12\pm 0.04$ was derived by 
\citet{brag06}, while \citet{sest06}
 report a slightly higher value [Fe/H]$=
0.02 \pm 0.04$: in 
both cases the metallicity is consistent with the solar value.

Given its age (we assume here 0.9\,Gyr), NGC\,3960 provides a good sample
to investigate the evolution of lithium abundance during the MS and,
in particular, to fill the gap in age coverage between the Hyades (0.6~Gyr)
and the $\sim 1.5-2.0$~Gyr clusters for which Li data are available
(NGC\,752, IC\,4651). 

\citet{sest05} investigated the timescales for Li depletion
during the MS for F and G--type stars,
by means of an homogeneous re-analysis of Li data for several open clusters.
They suggested that Li depletion is not a continuous process, but is
instead characterized by different timescales in different age intervals.
In particular, they showed that Li depletion slows down after the Hyades age
for stars in the temperature range $\sim 6050 - 6350$~K, while it stops
for cooler stars. Namely, for stars cooler
than $\sim 6050$~K, a plateau in Li abundance
is seen for ages older than $\sim 2$~Gyr; 
this age however represents an upper limit to the actual age when Li depletion
is not efficient any more. A more precise estimate of this age, which would
provide an useful constrain to models including extra-mixing during the main
sequence, requires Li data for cluster with ages between 0.6--2~Gyr.
We also mention in passing that since the study of \citet{sest05}  has been
used to infer the age of stars hosting extra-solar planets
\citep[e.g.][]{sozz07}, a finer sampling
of the 0.6--2~Gyr interval, would allow age dating these stars more accurately.

Besides allowing measurements of Li abundances, our Giraffe spectra
have been used to determine radial velocities and cluster membership
for the observed candidates.
This in turn will allow us to perform a revised analysis  
of contamination from field stars and cluster mass function.

Our paper is structured as follows: we describe in Section\,2 
the criterion adopted to select the spectroscopic targets and the observations
and in Section\,3 the analysis of the spectra, including the procedure used to
derive the radial velocity (RV) and the equivalent width (EW) of the Li line.
In Section\,4 we describe how we derive the effective temperatures and the Li
abundances of our targets. In Section\,5, we discuss the Li abundances for both
the cluster members and the field stars, by comparing our results with those of
stars of similar spectral type taken from literature; we also compare the
fraction of contaminating field stars in the spectroscopic sample with that
previously obtained statistically from photometric data. Our conclusions are
presented in Section\,7.

\section{Target selection and observations}
The targets observed with VLT/FLAMES were retrieved from the 
photometric catalog of \citet{pris04}, selecting objects in the 
cluster region, i.e. 
within 7\min\ from the cluster centroid. 
In total, we observed 113 candidate members
on the MS with  $16 \leq \rm V \leq 18$, corresponding
to spectral--types from F to early K:
stars with V larger than 18 are too cool and faint to 
reach a S/N larger than 40 that is required to have accurate measures of RVs 
and EWs; stars brighter than 16 are hotter than early F stars and therefore are
not good targets to investigate Li depletion in old main sequence stars.  

\begin{figure}
   \centering
   \includegraphics[width=9cm]{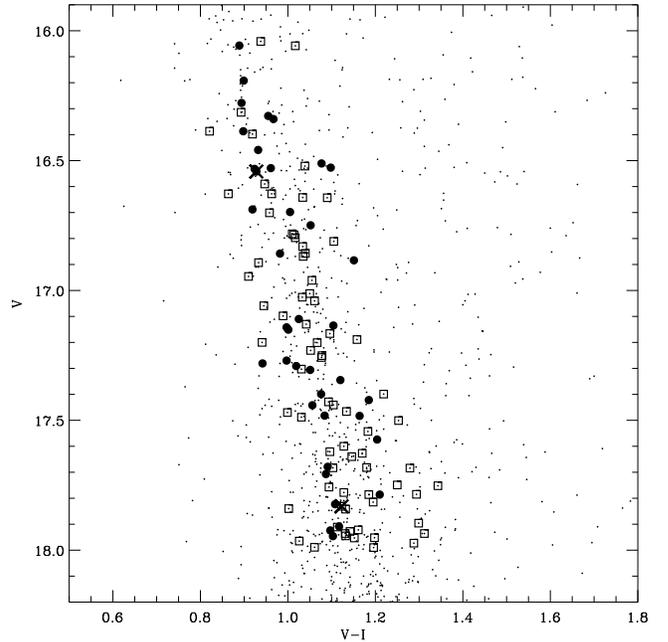}
   \caption{Color-magnitude diagram for the region within 7\min\ from the
   centroid of NGC\,3960 ( dots). Photometry is  from \citet{pris04} 
   and it is corrected
   for differential reddening. Targets observed with VLT/FLAMES are indicated
   with large symbols:  filled circles  are those with RV consistent with
   that of the cluster and  empty squares are non members having the RV
   different from the cluster members (see Section\,\ref{cluster_membership}).
   The two stars indicated with the {\sc X} symbols are candidate
   spectroscopic binaries (SB2).}
    \label{cmd_target} 
    \end{figure}

In Fig.\,\ref{cmd_target} we show
the V vs. V--I color-magnitude diagram 
for the region within 7\min\ from the
   centroid of NGC\,3960 ( dots) where photometry  is corrected
   for differential reddening. Targets observed with VLT/FLAMES are indicated
   with large symbols: filled circles   are those with RV consistent 
   with that of the cluster and  empty squares are non members having 
   the RV different from that of the cluster members, as we have found 
   with our analysis of RVs (see Section\,\ref{cluster_membership}).
	
The observations were carried out in Service Mode during February, March
and April 2004; the Log of the observations is provided in Table\,\ref{logbook}.
The cluster was covered by one configuration centered at 
 RA(2000)=11$^{\rm h}$50$^{\rm m}$31$^{\rm s}$\llap{.}400 
Dec(2000)=-55\deg41\min12\Sec80. Giraffe was used
in conjunction with the 316 lines/mm grating and order
sorting filters 14 (HR14) and 15 (HR15),  yielding  nominal 
resolving powers R=28,800 and 19,300, respectively. 
Spectral
coverages are from 630.8 to 670.1 for HR14 and from
660.7 to 696.5\,nm for HR15; they include H$\alpha$, the
Li\,{\sc i}\,670.8\,nm line and several features to be used for RV
 measurements. For each set-up, four 45\,min exposures were obtained.

\begin{figure}
   \centering
   \includegraphics{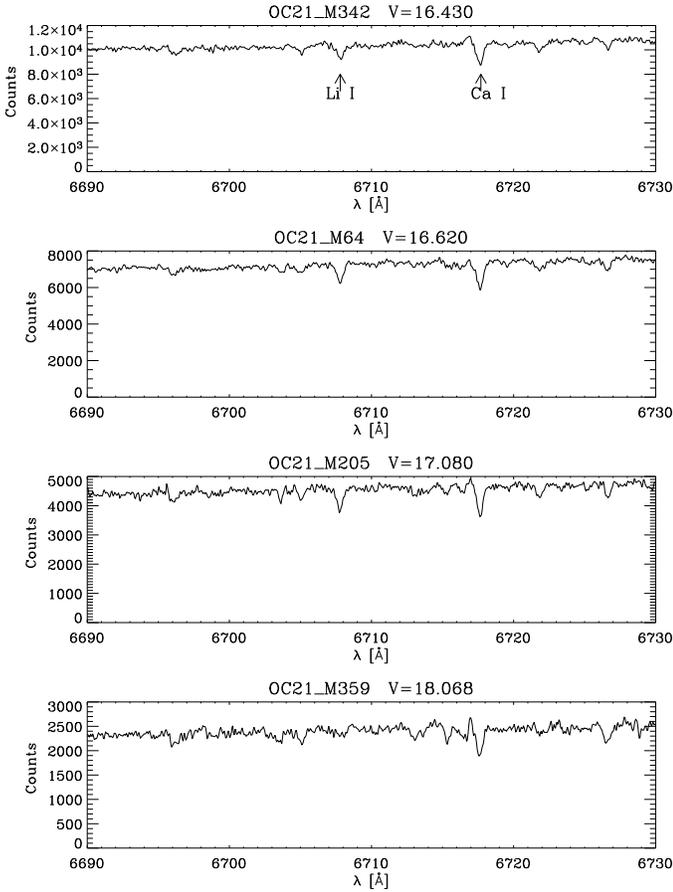}
   \caption{Examples of co-added sky-subtracted spectra in the spectral region
   that includes the Li line. The GIRAFFE ID given in Table\,\ref{table3}
   and the V magnitude of the stars are indicated.}
              \label{spectra} 
    \end{figure}

Data reduction was carried out using the GIRAFFE BLDRS
pipeline\footnote{version 1.0-- http://girbldrs.sourceforge.net/},
following the standard procedure and steps (Blecha et al. 2004).
Sky subtraction for the spectrum of each set-up and each exposure
was performed separately, namely by subtracting the median,
computed as in  \citet{jeff05},
of the 16 sky fiber spectra observed in the same exposure.
The four sky-subtracted spectra of each target obtained with the
HR15 set-up were then co-added, 
after applying the Doppler correction due to the different date of
observations. 
Examples of final co-added sky-subtracted spectra around the Li line
 are shown in 
Fig.\,\ref{spectra} where the GIRAFFE ID given in 
Table\,\ref{table3}
and the V magnitudes are indicated. S/N ratios  are $\sim$80
for the   stars \object{OC21-M342} and \object{OC21-M42}
 and $\sim$55 for \object{OC21-M205} and
 \object{OC21-M359}.
 Final S/N ratios range between 30 and 80, as estimated from the faintest and the
brightest stars in our sample.

\begin{table}
	\caption{Log-book of Giraffe/FLAMES observations.  Col.\,1 gives
	the observation date, col.\,2 gives the original name associated to each
	observing block, col.\,3 gives the exposure time and col.\,4 gives the
	grating used for each exposure.}
	\label{logbook}
\centering                          
\begin{tabular}{c c c c}        
\hline\hline 	
Obs. & Exp.   & Exp. Time &Grating\\
Date     &  Name      & (sec) & \\
\hline
04/03/2004 & M1-Li-a &2700&HR15\\
04/03/2004 & M2-Li-a &2700&HR15\\
04/03/2004 & M2-Li-c &2700&HR15\\
04/20/2004 & M1-ha   &2700&HR14\\
05/02/2004 & M1-Li-b &2700&HR15\\
05/02/2004 & M2-ha-a &2700&HR14\\
05/02/2004 & M2-ha-b &2700&HR14\\
05/20/2004 & M2-ha-c &2700&HR14\\
\hline                                   
\end{tabular}
\end{table}

\section{Data analysis}
\subsection{Radial velocities\label{radial_velocities}}
    Radial velocities (RV) in the heliocentric system 
    of the 8 sets of spectra were computed using the 
    function {\tt giCrossC} of the GIRAFFE 
    girBLDR pipeline. 
    We computed the cross-correlation function between all 
    the observed spectra and the synthetic template spectrum girKO 
    corresponding to a K0V star. 
    We used a delta RV limit to the cross-correlation window
    of 500\,km/s 
    in the wavelength ranges  [6310--6680]\,$\AA$, 
    for the spectra around the Li line, 
    and [6610--6810]\,$\AA$, 
    for the spectra around the H$\alpha$ line.  
    
        Final RVs for the stars of our sample were computed as 
    the mean value and the standard deviation
    of the RVs obtained with {\tt giCrossC} from the 8 different sets of 
    spectra.

    The RVs obtained using the function {\tt giCrossC}
    as described before were compared with those obtained by using
    a different template spectrum, girG2, that corresponds to a spectrum of a 
    solar-type star. The comparison does not evidence 
    significant differences. In addition, in
    order to check the reliability of the automatic cross-correlation performed
    by the pipeline, we computed the RV for 4 of the  8 sets of spectra using
    the IRAF \footnote{IRAF is distributed by the National Optical
Astronomical Observatories, which  are operated by the Association of
Universities for Research in Astronomy,  under contract with the National
Science Foundation.}
   task {\tt FXCOR} \citep{tonr79}; we used as template the spectrum
    of a relatively bright star ($V=16.5$, spectral type F8V)
    from our sample, that is considered a reliable cluster member, based on its
    RV as measured with {\tt giCrossC}. The relative RVs computed with 
    {\tt FXCOR} were converted into the heliocentric system using 
    the heliocentric
    RV of the template star computed with {\tt giCrossC}.
    The median of the differences  between the four sets of RVs obtained with 
    {\tt giCrossC} RVs minus {\tt FXCOR} RVs are 
0.13, --0.34, 0.12 and --1.05 Km/s  
    while the standard deviations are 2.13, 2.16,  1.81 and 2.10 Km/s.
    Such values indicate an excellent agreement 
    between the two methods. 
     The difference between the mean RVs  obtained with the two methods
    shows that only 8 stars
    (\object{OC21-M80}, \object{OC21-M273}, \object{OC21-M581},
     \object{OC21-M324}, \object{OC21-M282}, 
    \object{OC21-M185}, \object{OC21-M177}, \object{OC21-M41})
    have a difference in the mean RV obtained with the two methods larger 
    than 5\,km/s and only 2   (\object{OC21-M80} and \object{OC21-M282}) 
    have such difference larger than 10\,km/s.
    These objects do not show the Li line and   7  of them  
    are not cluster members since they have a RV 
    not consistent with that of the cluster 
    (see next section). The only star classified 
    as cluster member (\object{OC21-M273})  based on its
    RV does not show the Li line and therefore it is considered a
    contaminating field star.
     
        For 2 stars ( \object{OC21-M286}, \object{OC21-M136})
    the peak of the cross-correlation function performed with {\tt FXCOR},
    is not symmetric but shows a double peak, which indicates that these objects
    are double-lined spectroscopic binaries (SB2). 
    An inspection of the spectrum of these objects confirms the double spectral
    features, typical of SB2 stars.
    
    \subsubsection{Cluster membership\label{cluster_membership}}
\begin{figure}
   \centering
   \includegraphics[angle=-90,width=9cm]{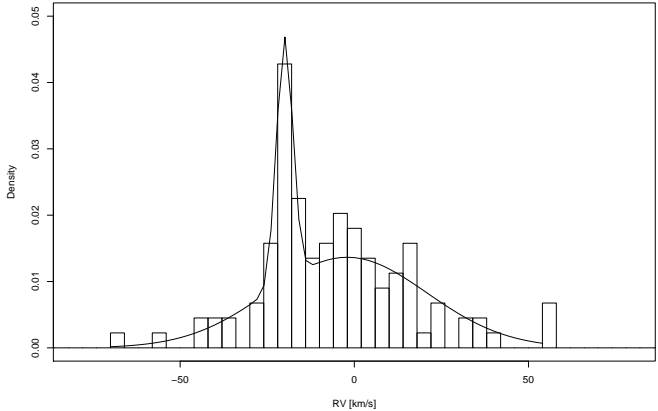}
   \caption{Density distribution of the RV of the whole sample that includes
   113 stars. The fitted double gaussian (see text) is indicated by the solid
   line.}
              \label{n3960_rv_med_paper} 
    \end{figure}

    The density distribution of the final RVs is shown in the histogram of  
    Fig.\,\ref{n3960_rv_med_paper} where a significant peak  around 
    RV=$-$20\,km/s
    indicates the presence of the cluster with respect to the RV distribution of
    the field stars which shows a secondary peak at about -2\,km/s.
     
    To derive the cluster membership based on the RV we fitted the final RV 
    distribution with a double gaussian using the ''maximum likelihood fitting",
    as in \citet{pris07}. We find that the cluster Gaussian  is centered on
    $-20.0\pm0.7$\,km/s with a standard deviation
    $\sigma=$2.3$\pm$0.6\,km/s,
    while the much broader
    field star RV distribution shows a peak at
    -2.1$\pm$2.6\,km/s, with a standard deviation of
    $\sigma=$22.9$\pm$-1.8\,km/s.
    The fitted curve is  indicated by the solid line in 
    Fig.\,\ref{n3960_rv_med_paper}.
    We note that the average cluster RV is smaller than the mean RV of NGC\,3960
    (equal to about
    -12\,km/s)  computed from  5  members studied in \citet{frie93}, 
    but in very good agreement with the value derived by \citet{sest06} from
    UVES spectra ($-22.6 \pm 0.9$\,km/s).
    
    The total number of
    possible cluster members within $\pm3\sigma$ of the cluster RV distribution
    is 39 including 16 contaminating field stars,
    as computed from the Gaussian distribution of the field stars.
    The 39 RV members are indicated by  filled circles in 
    Fig.\,\ref{cmd_target} together with the objects classified as non members,
    indicated by  empty squares, based on their RV.
	We note that some of the classified non members could be spectroscopic
	binaries of the cluster that cannot be easily distinguished from true
	non members. From an inspection of the single spectra, only 
	two objects, indicated by {\sc X} symbols in
	Fig\,\ref{cmd_target}, have been recognized from the characteristic
	double-line spectrum typical of SB2 stars 
	(see Section\,\ref{radial_velocities}).
	
Note that the identification of the cluster members does not allow us
to reduce the spread in the CMD since the differential reddening
corrections are  computed   using the reddening map shown in Fig.\,8 of
\citet{pris04} that gives the {\it relative} reddening values in subregions of
1\Min7$\times$1\Min7, with respect to the subregion where the cluster
centroid is located. They are therefore spatially-dependent corrections and are
 not calculated for each star individually. In addition,
  this statistical method includes also
the contaminating field stars.  
Individual spectral types estimates from
low resolution spectra are needed to derive the individual reddening and
therefore the accurate position of the members in the CMD. 
	 
\subsection{Li equivalent widths}
Li EWs were measured on the co-added spectra normalized 
to their continuum. The normalization was performed using the region of the
spectrum between 6693 and 6722\,$\AA$, that includes the \lithiumline\ line;
 we used the IRAF task {\tt CONTINUUM} with a second order Legendre function and a
 variable residual rejection limit chosen based on the visual inspection of the
 fitting result. The EWs of the \lithiumline\ line were measured using the IRAF 
 task {\tt SPLOT}, assuming a Gaussian profile, which is a good approximation
 since the EWs are all smaller than about 100\,m$\AA$ and therefore the line
 is not saturated.
 
 The continuum normalization and the EW measurements were repeated three times,
 in different dates; we take as final estimate of the EW
 the mean values of the three
 measures and as error the maximum  uncertainty.
 We detected the Li line in 55 objects, 29 with RV consistent with
membership and 26 RV non-members. Among  the remaining 58 stars,
we have   56 single objects and two SB2 binaries. For the  56 single stars
  we estimated an upper limit of the EW
 by measuring the EW  of the smallest line around  6707.8\,$\AA$. In some cases
 the spectra are of candidate M dwarf stars and show large molecular band around
 the Li line. For these stars, we considered the whole depression of the
 spectrum that includes the Li line and therefore the upper limits are very
 conservative.
 
 We corrected measured  EWs for the contribution of the Fe 
 {\footnotesize I} line at 6707.44\,$\AA$ using
 the relation EW(Fe)=20$(B-V)_0$-3m$\AA$, given in \citet{sode93a}
 for stars with solar metallicity, that is thus appropriate for our cluster
 \citep{sest06}.

\begin{figure}
   \centering
   \includegraphics[width=9cm]{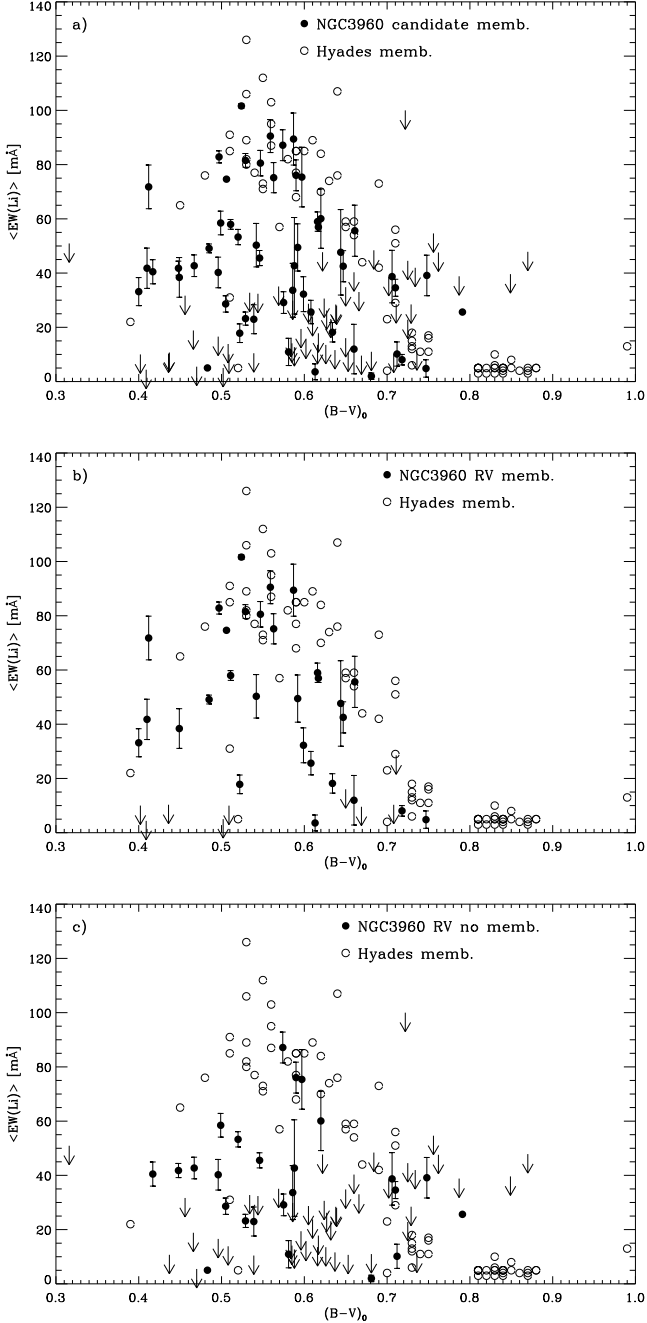}
   \caption{a) Li EWs and upper limits to EWs for all our candidate
   candidate members  as a function
 of the dereddened $(B-V)$ colors, corrected for differential reddening as 
 described in \citet{pris04}. Panels b) and c) show the Li EWs for the
 subsamples of the candidate members
 that are cluster members (panel b) or contaminating field stars (panel c),
 based on their RV. 
  Our measurements are compared with the EWs measured 
 for the stars in the Hyades \citep{sest05}.}
              \label{ew_vs_bv0} 
    \end{figure}

 Fig.\,\ref{ew_vs_bv0}a shows
 the Li EWs measured for all the 55 objects  where Li was detected
 as a function
 of the dereddened $(B-V)$ colors, corrected for differential reddening 
 as described
 in \citet{pris04}. 
 Fig.\,\ref{ew_vs_bv0}b shows the Li EWs for the
 stars which are cluster members based on their RV, while  
 Fig.\,\ref{ew_vs_bv0}c shows the Li EWs for the non members. 
 Upper limits are also plotted 
 in the three panels, as well as Li EWs measured 
 for the stars in the Hyades   \citep{sode90,thor93,sode95} .
 
 Whereas our measurements are characterized by a larger spread,
 panel b) indicates that the Li EWs of RV members 
are on average smaller than those of the Hyades members of similar color.
This is also true for most of the RV non members (panel c) with $(B-V)_0$
smaller than about 0.6 (F and G-type stars), while for redder colors,  Li
EWs are more comparable to those of Hyades stars.

\section{Li analysis}
\subsection{Effective temperatures}
\begin{figure}
   \centering
   \includegraphics[width=9cm]{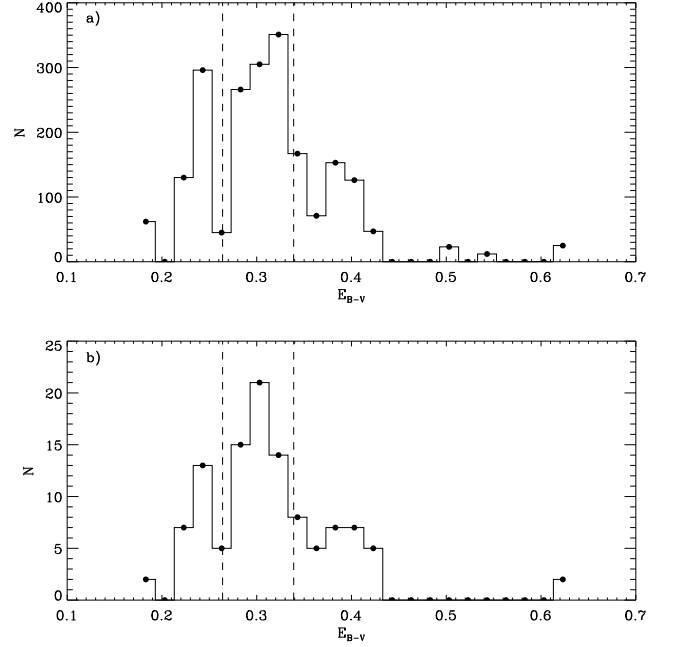}
   \caption{Reddening distribution computed as described in \citet{pris04}, 
   for all stars within 7\,arcmin from the cluster center (panel a) 
   and for the stars observed with Giraffe (panel b). 
   The  dashed lines indicate the 25$^{th}$ (0.26) and the 75$^{th}$ (0.34)
    percentiles of the two distributions (see text).}
              \label{diff_red} 
    \end{figure}

As in \citet{sest05}, effective temperatures were computed from the
dereddened $(B-V)$ colors, using the relation T$_{\rm
eff}=1800(B-V)_0^2-6103(B-V)_0+8899$ K, given by \citet{sode93b}.
Since photometric errors are smaller than 0.01  \,mag ,
errors in the effective temperatures are mainly due to the interstellar
reddening correction. Indeed, 
as discussed in \citet{pris04,brag06,bona06}, NGC\,3960
is affected by a relatively strong 
differential reddening. Fig.\,\ref{diff_red} shows the
reddening distribution computed as described in \citet{pris04}, for all stars
within 7 arcmin from the cluster center (panel a) and for the stars
observed with Giraffe (panel b). The  dashed lines indicate the
25$^{th}$ (0.26) and the 75$^{th}$ (0.34) percentiles of the two distributions 
 around the median value equal to 0.30;
we assume 
the semidifference of these two percentiles, equal to 0.04,
as the typical error of $E(B-V)$.
This range 
is in agreement with 0.29$\pm$0.05 found in
\citet{brag06} and with the range [0.03,0.34] given by \citet{bona06}. 

Assuming that $\sigma_{(B-V)}\sim\sigma_{E(B-V)}$,
i.e.  the error in the $(B-V)$ colors corrected for differential
reddening is of the order of the reddening error, we computed errors
 in the effective temperatures as the propagated uncertainties.

\subsection{Li abundances\label{li_abundances}}
Li abundances were computed from the EWs and the effective
temperatures by interpolating the growth curves of \citet{sode93a}.
Since these curves are computed assuming the local thermodynamic equilibrium 
(LTE), we corrected the  derived Li abundance, 
 for  non-local thermodynamic equilibrium using the \citet{carl94} code.

Errors in Li abundances were estimated by independently 
computing 
errors in Li abundance due to the effective temperature errors
and those due to EW errors. Finally, we  quadratically added these errors
to estimate our uncertainties in Li abundances.

\begin{figure}
   \centering
   \includegraphics[width=9cm]{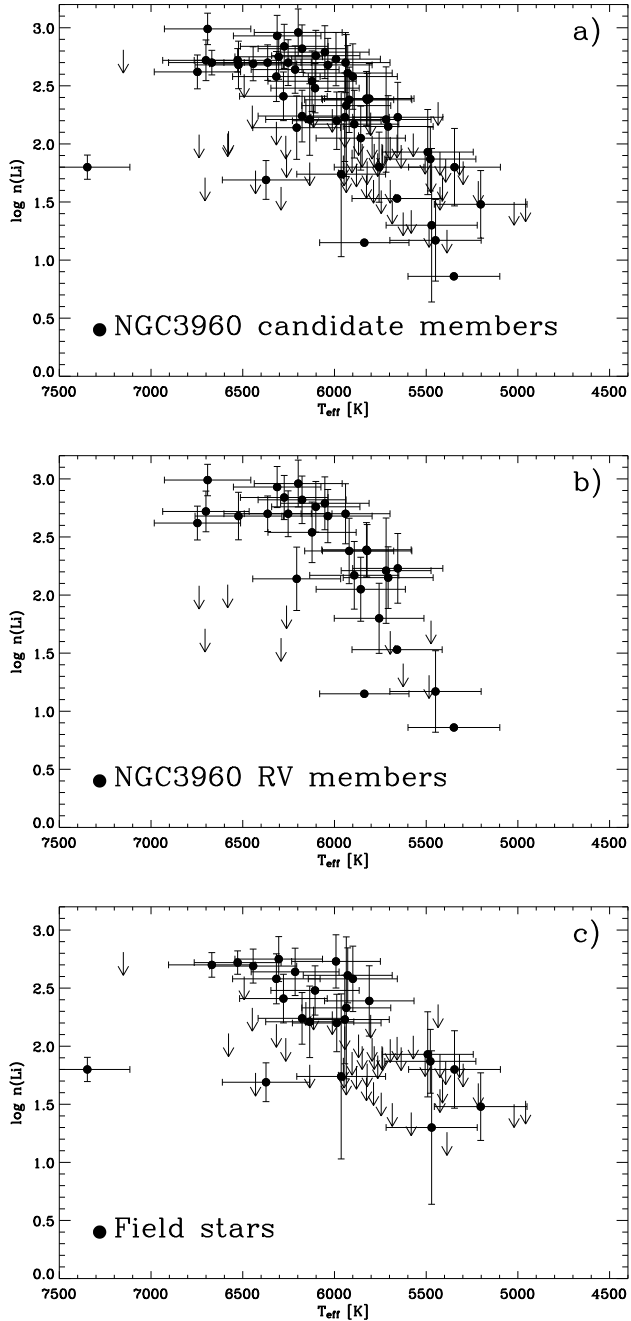}
   \caption{Li abundances derived from the effective temperatures computed using
   the $(B-V)_0$ colors corrected for differential reddening as described in
   \citet{pris04}. As in Fig.\,\ref{ew_vs_bv0}, the values are reported for the
   whole sample of photometric candidate members (panel a)
   and for the subsamples of RV members and non members (panels b) and c),
   respectively). Upper limits for the stars without Li are also indicated.}
              \label{litio_ab_NLTEcor} 
    \end{figure}

The computed Li abundances as a function of the temperatures
for all the photometric 
candidates are shown  in 
Fig.\,\ref{litio_ab_NLTEcor} (panel a);   
the RV members  and the non members are 
separately shown in panels b) and c), respectively. Upper limits are also
plotted.
Final data are given in Table\,\ref{table3} where   we list 
identification number of \citet{pris04},    Giraffe
identification name and   spectrum number,   celestial
coordinates,   V and B magnitudes, 
  V and B magnitudes corrected for differential reddening,
   membership flag 
based on the RV (0 means "non member",  1 means "member" and 2 means "binary"),
 effective temperatures computed from the (B-V)$_0$ colors, corrected for
differential reddening,    EW of the Li line and finally
   LTE and NLTE Li abundances. 
Note that in some cases, the NLTE Li abundances were not computed because the
LTE Li abundances were outside the range of allowed values for the \citet{carl94}
correction. Note also that for the 2 binaries, the EW of the Li line  were not
measured due to the complexity of the spectrum of these objects.
    
\begin{figure}
   \centering
   \includegraphics[width=9cm]{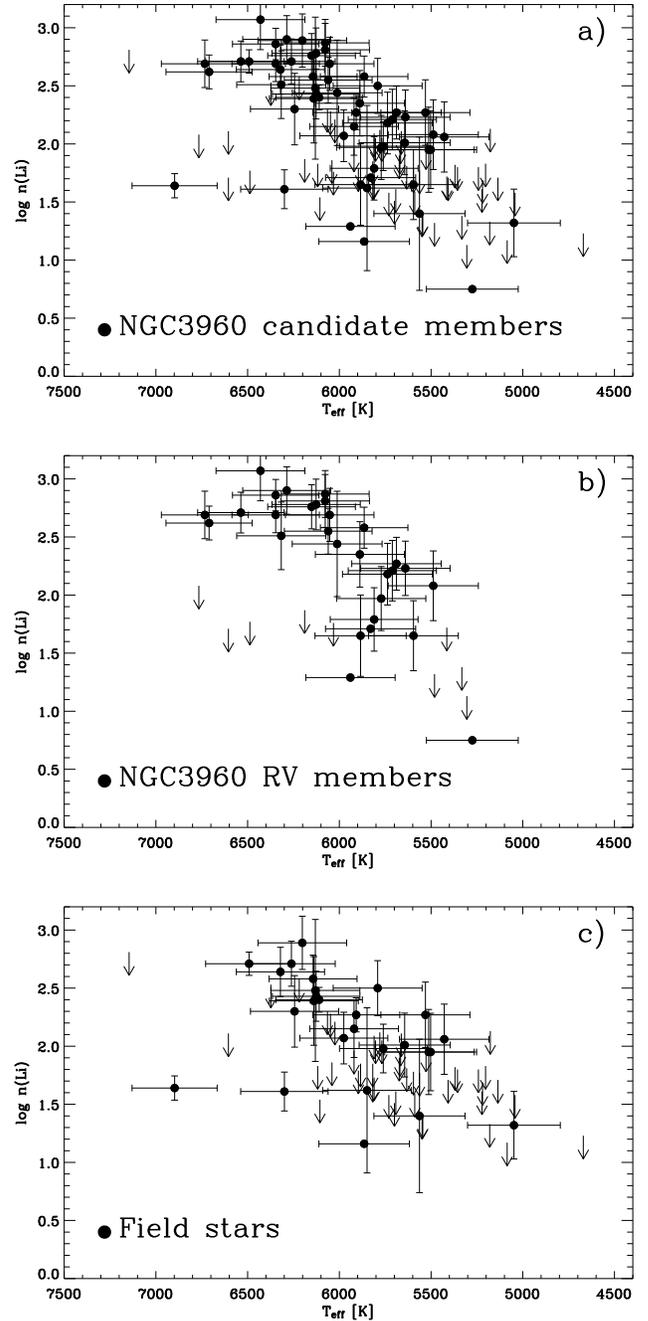}
   \caption{Li abundances derived from the effective temperatures computed using
   the  $(B-V)_0$ colors not corrected for differential reddening. The reported
   subsamples are the same described in Fig.\,\ref{litio_ab_NLTEcor}.}
              \label{litio_ab_NLTE} 
    \end{figure}

For comparison,  Fig.\,\ref{litio_ab_NLTE} shows the same plots of 
 Fig.\,\ref{litio_ab_NLTEcor}, but with the Li abundances derived from 
 the effective temperatures computed using
   the $(B-V)_0$ colors not corrected for differential reddening.

  We note that 
 Li abundances  in 
Fig.\,\ref{litio_ab_NLTE}b 
are slightly more spread out  than those in Fig.\,\ref{litio_ab_NLTEcor}b.
This suggests that temperatures obtained from colors corrected for differential
reddening are, on average, more accurate than those from uncorrected colors;
hence we will
adopt them for the following analysis.
However,   the adopted  reddening corrections are  derived
statistically from photometry assuming the distance of   
 the stars of NGC\,3960  and then the reddening correction
could not be appropriate for the field stars, especially if they are foreground
field stars.
Nevertheless, we find that the spread in panel c) of 
Fig.\,\ref{litio_ab_NLTEcor} is smaller
than that shown in panel c) of Fig.\,\ref{litio_ab_NLTE} and 
 similar to that found in \citet{pasq94} for field stars of
similar temperatures. 
    
\section{Discussion}
\subsection{Li abundances}
  Panel a) of Fig.\,\ref{litio_ab_NLTEcor} evidences a large spread in the Li 
abundance distribution:
 it includes both the coeval RV members of NGC\,3960 and the 
inhomogeneous sample of RV non members;  this is true
 especially for F-types and later spectral types,
  where the amount of the Li depletion
significantly depends on stellar age.
In the following sections we focus on the Li distribution for
cluster members and likely field stars.
\subsubsection{Cluster members}
 Filled circles in 
panel b) of Fig.\,\ref{litio_ab_NLTEcor}
shows the Li abundance distribution only for the 29 stars with detected Li
line and RV consistent with membership. As already mentioned in Section
\ref{radial_velocities}, we estimate that a total of 39 stars are RV members,
over a total of 113 objects studied in this work. Of the 39 objects, 16 are
 expected to
be contaminating field stars, with RV consistent with that of the cluster.
 
Based on Li only, we cannot definitively rule out
the possibility that the 10 stars with RV consistent
with membership and without Li are cluster members.
\footnote{The sample of the 10 stars includes
the binary with RV
consistent with that of the cluster, for which the EW has not been measured.}
On the one hand, for some of them we have inferred
upper limit values comparable or even higher than
measured Li EWs of stars with detected Li. On the other hand,
considering these 10 objects as cluster members would imply a large
spread in the Li abundances for warmer stars, which, again, we cannot
completely exclude. We note however that if these 10 stars or most
of them were indeed cluster members, we would have that most
of the contaminating field stars have high Li, which is rather
unlikely. On the contrary, 
we  make the more reasonable hypothesis that the 29 RV members with detected
Li are cluster members, while the 10 remaining objects
labeled as RV members, but without detected Li line
are among the 16 contaminating stars. 
This assumption would imply
that we have $\sim$60\% of non-members (10/16) without 
a measurable Li, that is in agreement with the expected 
fraction of field stars without Li (see next section).


 The remaining 6 contaminating
field stars show the Li line but cannot be identified in the sample of the 29
members shown in panel b). 

\begin{figure}
   \centering
   \includegraphics[width=9cm]{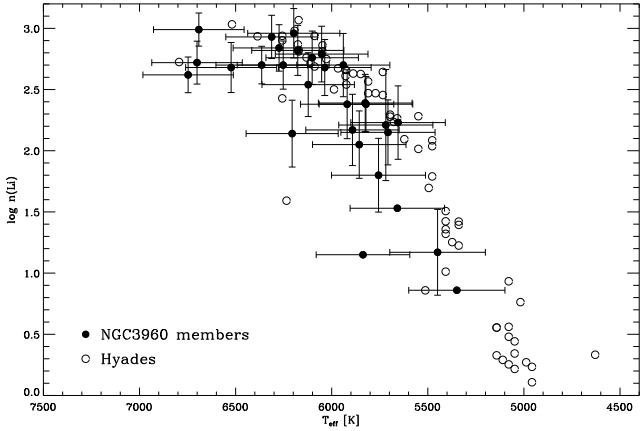}
   \caption{Li abundances of the 29 NGC\,3960 members (filled circles)
    as in Fig.\,\ref{litio_ab_NLTEcor} (panel b), compared with 
   the Hyades values (empty circles).}
              \label{comparison_hyades} 
    \end{figure}

In Fig.\,\ref{comparison_hyades}  we compare the Li abundances for the
29 NGC\,3960 candidate members
with the Li abundances  of the Hyades. The two distributions are almost
identical
for stars warmer than about 6000\,K, while the Li abundances of NGC\,3960 stars
cooler than 6000\,K are systematically smaller than those of Hyades
members of similar temperature. 
This result on the one hand supports our initial assumption
that NGC\,3960 is older than the Hyades; on the other hand, it allows
us to add a critical datapoint (see Sect.\,\ref{introdu}) on the empirical study
of the evolution of Li abundance with age. 

\begin{figure}
   \centering
   \includegraphics{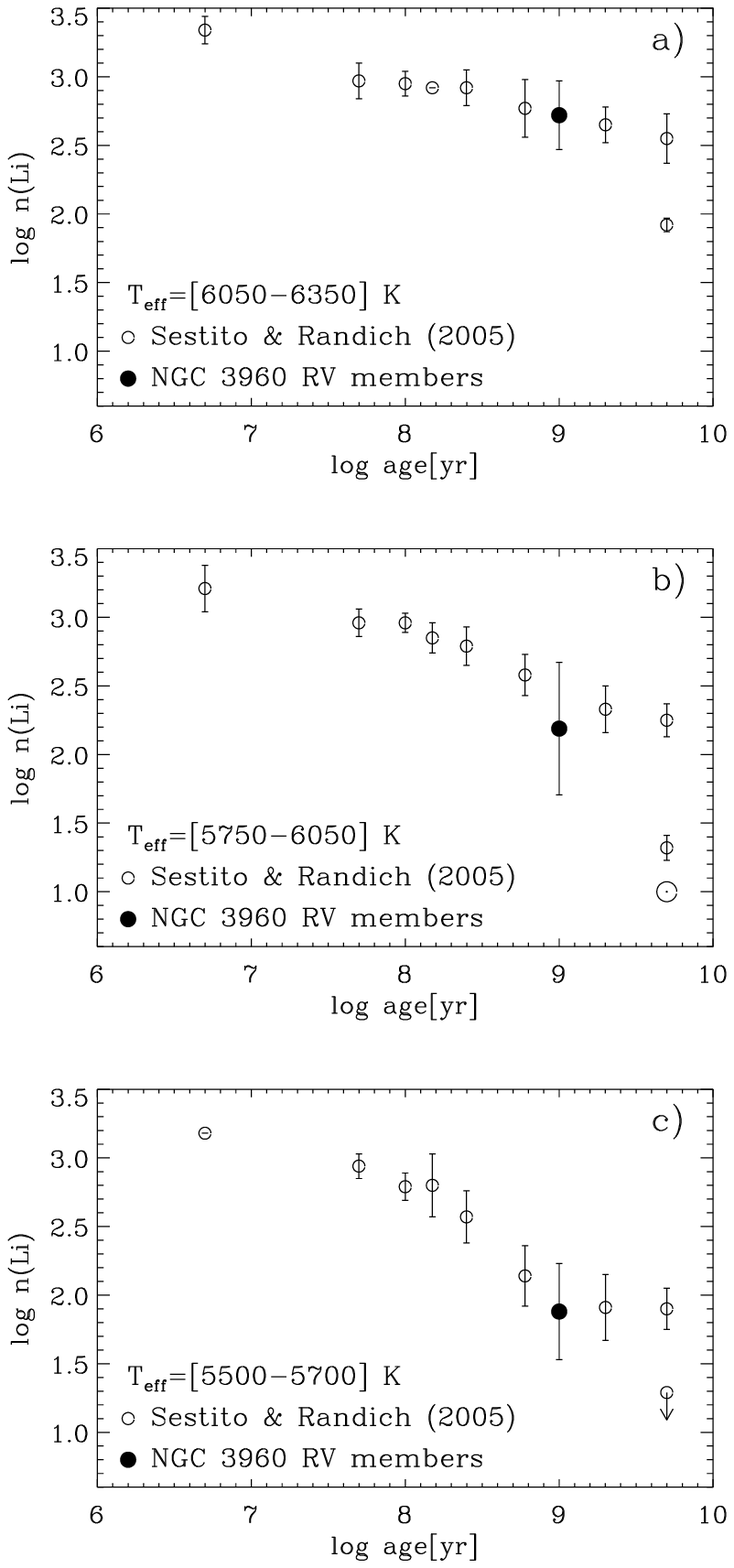}
   \caption{Plots adapted from Fig.\,7 of \citet{sest05} showing the 
   average log n(Li) as a function of cluster age, in three different
   temperature ranges.  Open circles are the data from \citet{sest05},
   while  filled circles are the values computed for NGC\,3960.}
              \label{comparison_sr05} 
    \end{figure}

Fig.\,\ref{comparison_sr05}, adapted from  \citet{sest05},
shows the average Li abundances as a function of  age, in three
different effective temperature ranges as computed by  \citet{sest05} 
for open clusters of different ages (open circles); 
the average values in the same temperature ranges derived here 
for NGC\,3960 using the 29 RV members with Li
 are indicated by filled circles
  and are given in Table\,\ref{average_Li}. 
  For comparison,
the position of the Sun is also indicated. 

\begin{table}
	\caption{Average of log n(Li) for NGC\,3960 in three different ranges
	of effective temperatures.}
	\label{average_Li}
\centering                          
\begin{tabular}{c c c c}        
\hline\hline 	
$\Delta$T$_{\rm eff}$ & $<$log n(Li)$>$   & error & error type\\
\hline
[6050--6350]\,K 	      & 2.72	      & 0.25	& $\sigma$\\

[5750--6050]\,K 	      & 2.19	      & 0.48	& $\sigma$\\

[5500--5700]\,K 	      & 1.88	      & 0.35	& maximum error\\
\hline                                   
\end{tabular}
\end{table}

For stars cooler than 5700\,K (panel c), the average Li abundance in NGC\,3960
is computed from only two stars and the associated error is the semidifference
of the two Li abundances. 

As discussed in \citet{sest05},
the upper panel shows that after a small amount of Li depletion 
occurring during the pre-main sequence (PMS) phase, no Li destruction
is present up to about 250\,Myr, while it restarts afterwards.
The average abundances for NGC\,3960 confirm the trend of slow
(but present) depletion at ages older than the Hyades.

Panels b) and c) of Fig.\,\ref{comparison_sr05} 
indicate that for cooler stars, after the phase
of PMS Li destruction, 
Li depletion restarts significantly after about 200\,Myr. Most important,
our analysis allows us to put tighter constraints on the age
at which Li depletion is not present any more; specifically
the average Li
abundances derived for NGC\,3960 shows that the Li plateau  
might start already
at $\sim$ 1\,Gyr rather than 2\,Gyr: the average abundance of NGC~3960
is indeed closer to that of the 2~Gyr clusters, rather than to the Hyades,
although, due to the large $\sigma$,~it is not completely
inconsistent with the latter. Also,
average Li abundances in NGC\,3960 could be lowered by the inclusion of a few
non members with RV consistent with that of NGC\,3960 but which might
have lower Li abundances.
For example, the average Li abundance of stars in panel b) has been
computed including in the sample
the star with log\,n(Li)$\simeq$1.15 and
T$_{\rm eff}\simeq 5800$\,K; its Li abundance is significantly deviating
from the mean Li pattern, suggesting that this object might come from the
sample of contaminating field stars. By excluding this object from
the sample we would get a smaller $\sigma$ (0.31) and a slightly higher average
(2.32) that is however consistent with that of older clusters.

We stress that our analysis has been carried out consistently with
that of \citet{sest05} and thus the comparison of the average abundances
should not be affected by systematic errors. 

As discussed by \citet{sest05},
none of the models/mechanisms proposed to explain the occurrence
of MS Li depletion is also able to reproduce the plateau, since all of them
predict that Li depletion should continue at old ages.
To our knowledge, no new models predicting the existence of the
plateau have been presented. Our study, not only reinforces 
the empirical evidence for the plateau and thus the need for such models, but
also provides an additional constrain to be taken into account.
\subsubsection{Field stars}
Filled circles in panel c) of Fig.\,\ref{litio_ab_NLTEcor}
shows the Li abundance distribution for the 26 objects that are contaminating
field stars according to their RV, but do show the Li line.
Their distribution is not remarkably different from that of NGC\,3960 
members,
although it is characterized by a much larger dispersion. 
As discussed in \citet{pris04}, the differential reddening correction was
computed as the distance (along the reddening vector) 
of the position in the CMD of each  star from the assumed main sequence at 
the cluster distance. Therefore 
such correction does not take into account the distance 
spread  from the Sun for field stars and this explains the  large spread 
in the Li abundance.

Nevertheless, several
stars with high abundance ($\log$~n(Li) $> 2$) are present.
More specifically,    
 we find that the field 
stars  hotter
than about 6400\,K (early F-type) show a Li abundance of about 2.7,
consistent with that found for stars of similar temperatures for NGC\,3960;
stars cooler than 6400\,K and hotter than 5700\,K (late F-type and early G-type)
show a large spread in the Li abundance with values similar or smaller than 
those found for the same temperature range for NGC\,3960; finally, stars
with temperatures smaller than about 5500\,K have Li abundances somewhat
larger than
those found for NGC\,3960. Note however that effective temperatures (and thus
Li abundances) for these stars might have been overestimated,
 since we have 
assumed for them the same reddening as for NGC\,3960 members.
Empty squares in 
panel c) of Fig.\,\ref{litio_ab_NLTE}
indicate Li abundances computed assuming E(B-V)=0 and
no differential reddening correction.
As expected they have temperatures much cooler than those obtained assuming the
cluster reddening and consequently, based on the
growth curves of \citet{sode93a}, smaller Li abundances.

We exclude that the stars in panel c)  are binary cluster members (SB1) 
with discrepant RV due to the orbital component, since for
these objects, the
distribution of the difference between the maximum and the minimum values of
the RV measured from our spectra (acquired within about 
one month and half, see Table\,\ref{logbook})
is very similar to that obtained for the sample of RV members.
Therefore we conclude that
they are, most likely,  a sample of field stars with Li.

If we consider that we have 74 non members for RV (i.e. 113-39), including the
binary,  
plus 16 contaminating field stars
stars with RV consistent with that of NGC\,3960, we have a total of 
90 field
stars; those with the Li line are 26 within the sample of RV non
members, plus the remaining 6 contaminating field stars in the 
sample of RV members.
Therefore, the fraction of field stars with the Li line is (26+6)/90, i.e.
about one third of the sample. This fraction is slightly lower than that
found by \citet{pasq94} who found about one half of the G dwarfs analyzed in
their work with high 
Li content (2.0$\lesssim$log\,n(Li)$\lesssim$3.0)
and apparently old age. Also note that all solar-type stars in
the very old cluster NGC\,188 ($\sim 6$~Gyr) with available Li measurements
have abundances above 2.0 \citep{rand03}, i.e.  a
relatively high Li is not inconsistent
with a very old age.

In summary, the presence of several Li-rich F and
G-type field stars in our  sample is not surprising. 
The large spread in their Li abundance is both
consistent with a mixed population of stars older or similar to NGC\,3960,
and also possibly/in part due to the fact that, in addition to stellar age,
another unknown parameter can regulate MS Li depletion 
\citep[e.g.][and references therein]{pasq94,pins97,pasq97,char05,rand06}.
%

\subsection{Membership and contamination}
Table\,\ref{membership_tab} gives the total number of stars 
defined as cluster members or non members based on the RV and the
presence or not of the Li line in their spectrum.
As already discussed in the previous section, we find a total of 29
objects that satisfy both membership criteria. This sample contains
6 of the 16 contaminating stars with RV consistent with that of NGC\,3960,
estimated by the RV distribution of the field stars 
(see Section\,\ref{radial_velocities}). The remaining 10 contaminating stars
can be individually
distinguished since do not show the Li line. If we add the 74 stars with RV
not consistent with the cluster membership, we have a total of 84 individually
known non members
over a total of 113 observed stars.

\begin{figure}[!ht]
   \centering
   \includegraphics[width=9cm]{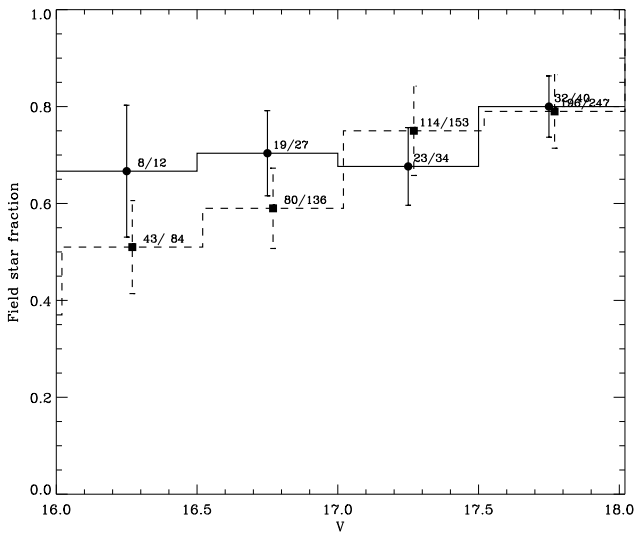}
   \caption{Fraction of contaminating field stars estimated 
   in this work from individual membership derived by RV and the Li
   line ( solid line histogram) and in \citet{pris04} 
   from a statistical analysis of a field region around NGC\,3960
   ( dashed line histogram).}
              \label{contamination} 
    \end{figure}

We considered these 84 objects to estimate the fraction of
 contaminating field stars  in our sample as a function of the V magnitude. 
 The results are shown by the
 solid line histogram plotted in  
Fig.\,\ref{contamination}; 
 the number of contaminating stars and the total number of objects 
 within the four magnitude ranges are also indicated; 
 error bars were computed from the binomial distribution.
 The resulting fractions are compared with
 the analogous values derived in \citet{pris04}     
  where the contamination was statistically
 derived using a field region outside the cluster region 
 (dashed line histogram); in this case,
  errors were computed from the propagation
 of the poisson errors on the number of field stars and of the total stars.
 The comparison shows that, within the  errors, the fractions of contaminating 
 stars derived with the two methods are compatible. 
 
We note, however, that, among the 84 contaminating stars, 
the sample of 74 non members for the RV could include a {\it small} 
fraction of binaries of NGC\,3960 and therefore these numbers could be slightly 
overestimated. On the other hand, we did not include in the field star sample
 the 6 contaminating stars with RV consistent with that of the cluster and with the Li line.
 Nevertheless, since these effects should be within the error bars, we conclude that the
 fraction of contaminating field stars within the spectroscopic sample studied in this work
 is comparable with that estimated in \citet{pris04} with photometric data.
 We note that the comparison is consistent since the stars observed in this spectroscopic work
  have been randomly selected from
 the list of photometric candidate members given in \citet{pris04}. 
 This result allows us to confirm the Mass Function derived in \citet{pris04} that strongly
 depends on  the correction for the field star contamination.

\begin{table}
\centering
\tabcolsep 0.1truecm
\caption{Number of  stars
 with  RV consistent or not (within $3\sigma$) with the cluster membership
  (indicated by Y or N) and with or without 
Li (indicated by Y or N).
 The total number for each criterion is also given.
The two binaries are specifically indicated in the given samples.}
\vspace{0.5cm}
\begin{tabular}{|l||*{4}{c|}}\hline
\backslashbox{Li}{RV}
&\makebox[3em]{Y}&\makebox[3em]{N}&\makebox[3em]{Tot.}\\\hline\hline
Y       & 29	  & 26	    & 55	\\\hline
N       & 10(9+1) 	  & 48(47+1)    & 58(56+2)	\\\hline
Tot.    & 39(38+1)      & 74(73+1)      & 113(111+2)	\\\hline
\end{tabular}
\label{membership_tab}
\end{table}

\section{Summary and Conclusions}
We used VLT/FLAMES Giraffe spectra to determine the radial velocities and
thus the membership of a
sample of 113 photometrically selected 
candidate cluster members. We find that the average cluster radial velocity is 
$-20.0 \pm 0.7$\,km/s with a standard deviation
$\sigma=$2.3$\pm$0.6\,km/s. As expected from the high fraction of 
  contaminating stars, only 39 objects have RV consistent with that of the
  cluster. Among these, 16 are expected to be contaminating field stars and 10
  of these have been identified since do not show the Li line.  
We find that the spread in the CMD for the sample of cluster members is not
reduced since the adopted reddening correction is statistically  
reliable. Individual spectral types of such objects are  necessary to
accurately derive their temperatures and the position in the CMD.
    
From the analysis of the Li line 
we derived Li abundances for cluster and field stars. We find that by using
photometry corrected for differential reddening, Li abundance distribution
of the cluster members shows a spread smaller than that found using uncorrected
photometry.  

The average Li abundances computed  
 for stars in three different temperature ranges confirms the trend,
   already found in \citet{sest05}, of slow
 (but present) depletion at ages older than the Hyades for stars with
 temperature larger than about 6000\,K. 
 For cooler stars, the inclusion of the average Li abundance of NGC\,3960 
in the distribution of Li abundance as a function of the age, allows us to
conclude that the age at which Li depletion is not present any more is  
 very likely closer to
1\,Gyr rather than 2\,Gyr as inferred by \citet{sest05}, on the base of the
incomplete sample of open clusters with ages older than 2\,Gyr and younger 
than the Hyades.

The fraction of field stars with an evident Li line is about one third, while
the fraction of field stars, as a function of the magnitude, is consistent with
 that derived statistically from photometry. This allows us to confirm the
 conclusions about the Mass Function derived in \citet{pris04} that strongly
 depends on the estimate of the field star contamination. 
\begin{acknowledgements}
We thank the ESO Paranal staff for performing the
service mode observations. This work has made extensive use of the
services of WEBDA, ADS, CDS etc. The research presented here
has been supported by an INAF grant on
{\it Stellar clusters as probes of star formation and early stellar 
evolution (PI: F.Palla)}. 
L.P. thanks   Paolo Span\`o for his contribution to the RV analysis and
Giusi Micela for useful discussions on this work.
      
\end{acknowledgements}
\bibliographystyle{aa}
\bibliography{n3960}
\input{table}
\end{document}

%% file: table.tex
\longtabL{4}{
\begin{landscape}
\tabcolsep 0.12truecm
\renewcommand{\footnoterule}{}  
\begin{longtable}{ccccccccccccccc}
\caption{\label{table3} Stellar parameters for the  NGC\,3960 
stars observed with GIRAFFE. 
{\bf Col.\,1 gives the
identification number of \citet{pris04}, Cols.\,2 and 3 give the Giraffe
identification name and the spectrum number, Cols.\,4 and 5 give the celestial
coordinates, Cols.\,6 and 7 give the V and B magnitudes while Cols.\,8 and 9 
give the V and B magnitudes corrected for differential reddening, Col.\,11 
gives the membership flag 
based on the RV (0 means "non member",  1 means "member" and 2 means "binary");
 Col.\,12 gives the
effective temperatures computed from the (B-V)$_0$ colors, corrected for
differential reddening; Col.\,13 gives the EW of the Li line and finally
Col.\,14 and 15 give the LTE and NLTE Li abundances, respectively.}}\\
\hline\hline
   ID &         ID &  Sp &     RA(2000) &    Dec(2000) &      V &      B &   Vcor &   Bcor &     RV &   M. & Teff &EW(Li) &log n(Li) &log n(Li) \\
  P04 &    Giraffe &     &        [deg] &        [deg] &        &        &        &        & [km/s] &RV &  [K]
  &m$\AA$ &      LTE &     NLTE \\
\hline
\endfirsthead
\caption{continued.}\\
\hline\hline
   ID &         ID &  Sp &     RA(2000) &    Dec(2000) &      V &      B &   Vcor &   Bcor &     RV &   M. & Teff &EW(Li) &log n(Li) &log n(Li) \\
  P04 &    Giraffe &     &        [deg] &        [deg] &        &        &        &        & [km/s] &RV &  [K] &m$\AA$ &      LTE &     NLTE \\
\hline
\endhead
\hline
\endfoot
 3983    & \object{  OC21-M104}&   2 & 177.61642456 & -55.64477921 & 16.594$\pm$  0.003 & 17.514$\pm$  0.006 & 16.527 & 17.425 & -16.85$\pm$   1.60 & 1 & 5771$\pm$ 243 &  34$\pm$ 4 & 2.04 &     2.05$\pm$     0.28 \\
 3185    & \object{   OC21-M80}&   3 & 177.64501953 & -55.66764832 & 16.388$\pm$  0.004 & 16.995$\pm$  0.003 & 16.387 & 16.993 &  -1.95$\pm$   5.21 & 0 & 7146$\pm$ 232 & $<$   50   &$<$ 2.89 &$<$     2.81               \\
 4873    & \object{  OC21-M301}&   4 & 177.62677002 & -55.60786819 & 17.435$\pm$  0.003 & 18.420$\pm$  0.005 & 17.130 & 18.017 & -36.28$\pm$   0.91 & 0 & 5530$\pm$ 242 &  84$\pm$11 & 2.61 &     2.58$\pm$     0.28 \\
 4850    & \object{  OC21-M299}&   5 & 177.61761475 & -55.61310959 & 16.645$\pm$  0.003 & 17.555$\pm$  0.006 & 16.340 & 17.152 & -15.42$\pm$   0.67 & 1 & 5810$\pm$ 239 &  25$\pm$ 3 & 2.17 &     2.14$\pm$     0.27 \\
 5446    & \object{  OC21-M339}&   6 & 177.63058472 & -55.60300064 & 16.935$\pm$  0.003 & 17.771$\pm$  0.003 & 16.701 & 17.461 &  -4.28$\pm$   0.99 & 0 & 6105$\pm$ 237 & $<$    5   &$<$ 1.81 &$<$     1.77               \\
 5410    & \object{  OC21-M566}&   7 & 177.53910828 & -55.58911514 & 17.900$\pm$  0.005 & 19.030$\pm$  0.006 & 17.749 & 18.830 &   2.16$\pm$   0.70 & 0 & 5048$\pm$ 252 &  38$\pm$ 0 & 1.36 &     1.48$\pm$     0.29 \\
 5443    & \object{  OC21-M571}&   8 & 177.57386780 & -55.57823944 & 17.830$\pm$  0.004 & 18.785$\pm$  0.005 & 17.679 & 18.585 & -20.03$\pm$   1.10 & 1 & 5640$\pm$ 243 &  68$\pm$ 3 & 2.40 &     2.39$\pm$     0.23 \\
 3867    & \object{  OC21-M260}&   9 & 177.58436584 & -55.64488983 & 17.454$\pm$  0.003 & 18.284$\pm$  0.007 & 17.600 & 18.478 & -42.39$\pm$   3.37 & 0 & 6130$\pm$ 242 &  51$\pm$17 & 2.34 &     2.33$\pm$     0.61 \\
 3055    & \object{   OC21-M65}&  10 & 177.63095093 & -55.67090225 & 16.949$\pm$  0.002 & 17.792$\pm$  0.003 & 16.858 & 17.672 & -19.58$\pm$   1.53 & 1 & 6076$\pm$ 239 & 109$\pm$ 0 & 3.06 &     2.96$\pm$     0.20 \\
 4011    & \object{  OC21-M273}&  11 & 177.59687805 & -55.63790131 & 16.345$\pm$  0.002 & 17.066$\pm$  0.004 & 16.278 & 16.977 & -15.49$\pm$   7.73 & 1 & 6604$\pm$ 235 & $<$    4   &$<$ 1.76 &$<$     1.71               \\
 3044    & \object{   OC21-M64}&  12 & 177.60371399 & -55.67435837 & 16.620$\pm$  0.002 & 17.445$\pm$  0.005 & 16.529 & 17.325 & -20.82$\pm$   0.44 & 1 & 6151$\pm$ 238 &  81$\pm$ 0 & 2.92 &     2.84$\pm$     0.19 \\
 4672    & \object{  OC21-M540}&  13 & 177.51086426 & -55.60569763 & 17.953$\pm$  0.004 & 18.901$\pm$  0.007 & 17.928 & 18.868 &  15.99$\pm$   3.48 & 0 & 5666$\pm$ 245 & $<$   34   &$<$ 2.03 &$<$     2.06               \\
 5551    & \object{  OC21-M340}&  14 & 177.66407776 & -55.59788513 & 17.095$\pm$  0.003 & 18.064$\pm$  0.004 & 16.857 & 17.749 &  13.11$\pm$  10.88 & 0 & 5588$\pm$ 242 & $<$   15   &$<$ 1.82 &$<$     1.83               \\
 4060    & \object{  OC21-M108}&  15 & 177.64192200 & -55.65436554 & 16.659$\pm$  0.003 & 17.458$\pm$  0.004 & 16.628 & 17.417 & -40.03$\pm$   1.94 & 0 & 6260$\pm$ 238 &  65$\pm$ 4 & 2.81 &     2.75$\pm$     0.19 \\
 3282    & \object{   OC21-M88}&  16 & 177.71395874 & -55.67269516 & 17.806$\pm$  0.007 & 18.711$\pm$  0.006 & 17.946 & 18.896 & -20.29$\pm$   0.98 & 1 & 5829$\pm$ 245 &  22$\pm$ 9 & 1.49 &     1.53$\pm$      --\\
 5464    & \object{  OC21-M572}&  17 & 177.63177490 & -55.59671402 & 17.675$\pm$  0.004 & 18.651$\pm$  0.004 & 17.441 & 18.341 &   9.99$\pm$   1.55 & 0 & 5563$\pm$ 243 & $<$   23   &$<$ 1.99 &$<$     2.00               \\
 5368    & \object{  OC21-M564}&  18 & 177.58049011 & -55.60188675 & 16.933$\pm$  0.004 & 17.801$\pm$  0.005 & 16.782 & 17.601 &  -8.75$\pm$   0.52 & 0 & 5975$\pm$ 239 &  30$\pm$ 2 & 2.27 &     2.24$\pm$     0.22 \\
 5566    & \object{  OC21-M581}&  19 & 177.64627075 & -55.59479523 & 17.922$\pm$  0.004 & 19.011$\pm$  0.007 & 17.684 & 18.696 &  -1.25$\pm$   6.67 & 0 & 5176$\pm$ 248 & $<$  100   &$<$ 2.33 &$<$     2.36               \\
 5893    & \object{  OC21-M607}&  20 & 177.59971619 & -55.57306290 & 16.706$\pm$  0.002 & 17.624$\pm$  0.003 & 16.521 & 17.380 &   4.67$\pm$   0.69 & 0 & 5779$\pm$ 241 & $<$   35   &$<$ 2.32 &$<$     2.30               \\
 4256    & \object{  OC21-M125}&  21 & 177.69602966 & -55.65552521 & 17.305$\pm$  0.004 & 18.381$\pm$  0.006 & 17.501 & 18.640 &  -0.97$\pm$   0.71 & 0 & 5219$\pm$ 256 & $<$   39   &$<$ 1.35 &$<$     1.50               \\
 3969    & \object{  OC21-M102}&  22 & 177.61032104 & -55.64798737 & 17.641$\pm$  0.003 & 18.700$\pm$  0.009 & 17.574 & 18.611 & -20.30$\pm$   0.77 & 1 & 5274$\pm$ 250 &  16$\pm$ 3 & 0.77 &     0.86$\pm$      -- \\
 5095    & \object{  OC21-M327}&  23 & 177.74018860 & -55.60593796 & 18.074$\pm$  0.004 & 19.338$\pm$  0.008 & 17.752 & 18.912 &  55.82$\pm$   1.38 & 0 & 4669$\pm$ 257 & $<$   47   &$<$ 1.36 &$<$     1.53               \\
 5078    & \object{  OC21-M324}&  24 & 177.71362305 & -55.60957336 & 16.720$\pm$  0.002 & 17.570$\pm$  0.003 & 16.398 & 17.144 &  31.76$\pm$   8.73 & 0 & 6048$\pm$ 237 & $<$   31   &$<$ 2.66 &$<$     2.60               \\
 5647    & \object{  OC21-M344}&  25 & 177.71800232 & -55.60044479 & 18.321$\pm$  0.006 & 19.396$\pm$  0.027 & 17.953 & 18.909 & -11.88$\pm$   6.99 & 0 & 5222$\pm$ 246 & $<$   33   &$<$ 1.95 &$<$     1.99               \\
 5098    & \object{  OC21-M328}&  26 & 177.70584106 & -55.60464096 & 17.020$\pm$  0.002 & 17.956$\pm$  0.003 & 16.698 & 17.530 & -20.18$\pm$   0.79 & 1 & 5710$\pm$ 240 &  58$\pm$ 8 & 2.58 &     2.54$\pm$     0.26 \\
 5072    & \object{  OC21-M323}&  28 & 177.73262024 & -55.61132050 & 17.810$\pm$  0.006 & 18.824$\pm$  0.006 & 17.488 & 18.398 &  -4.05$\pm$   0.74 & 0 & 5428$\pm$ 243 &  69$\pm$10 & 2.40 &     2.39$\pm$     0.30 \\
 3217    & \object{   OC21-M83}&  29 & 177.66282654 & -55.65973282 & 17.232$\pm$  0.003 & 18.140$\pm$  0.004 & 17.231 & 18.138 &  36.26$\pm$   1.15 & 0 & 5817$\pm$ 243 & $<$   17   &$<$ 1.83 &$<$     1.85               \\
 5569    & \object{  OC21-M342}&  30 & 177.67813110 & -55.59438324 & 16.430$\pm$  0.003 & 17.209$\pm$  0.018 & 16.192 & 16.894 & -24.19$\pm$   1.07 & 1 & 6346$\pm$ 235 &  77$\pm$ 8 & 3.08 &     2.99$\pm$     0.14 \\
 6002    & \object{  OC21-M618}&  31 & 177.69937134 & -55.56566238 & 17.618$\pm$  0.005 & 18.514$\pm$  0.007 & 17.281 & 18.068 & -20.67$\pm$   1.75 & 1 & 5864$\pm$ 238 &  89$\pm$ 2 & 3.02 &     2.93$\pm$     0.18 \\
 4425    & \object{  OC21-M282}&  33 & 177.79666138 & -55.64904404 & 17.703$\pm$  0.006 & 18.652$\pm$  0.004 & 17.936 & 18.960 &  -5.83$\pm$   2.56 & 0 & 5662$\pm$ 249 & $<$   41   &$<$ 1.80 &$<$     1.87               \\
 4502    & \object{  OC21-M286}&  34 & 177.79386902 & -55.64278793 & 17.598$\pm$  0.030 & 18.420$\pm$  0.006 & 17.831 & 18.728 &  -2.50$\pm$   3.12 & 2 & 6163$\pm$ 243 &            &         &                           \\
 4484    & \object{  OC21-M284}&  36 & 177.79835510 & -55.63433838 & 16.956$\pm$  0.002 & 17.852$\pm$  0.003 & 17.189 & 18.160 &   7.00$\pm$   0.99 & 0 & 5864$\pm$ 246 &  12$\pm$ 1 & 0.62 &      -- \\
 4350    & \object{  OC21-M279}&  37 & 177.77566528 & -55.63901520 & 17.032$\pm$  0.002 & 17.888$\pm$  0.003 & 17.258 & 18.186 &  15.58$\pm$   0.56 & 0 & 6023$\pm$ 244 & $<$   28   &$<$ 1.98 &$<$     2.00               \\
 5124    & \object{  OC21-M331}&  38 & 177.77940369 & -55.62976456 & 16.948$\pm$  0.004 & 17.738$\pm$  0.006 & 16.893 & 17.666 & -43.74$\pm$   1.48 & 0 & 6299$\pm$ 238 &  11$\pm$ 0 & 1.72 &     1.69$\pm$     0.17 \\
 3133    & \object{   OC21-M76}&  39 & 177.64811707 & -55.68086243 & 17.423$\pm$  0.004 & 18.422$\pm$  0.012 & 17.422 & 18.420 & -19.59$\pm$   0.64 & 1 & 5480$\pm$ 248 & $<$   10   &$<$ 1.25 &$<$     1.31               \\
 5156    & \object{  OC21-M336}&  41 & 177.76028442 & -55.62010193 & 18.044$\pm$  0.005 & 18.935$\pm$  0.007 & 17.989 & 18.863 & -39.11$\pm$   1.93 & 0 & 5884$\pm$ 242 & $<$   16   &$<$ 1.90 &$<$     1.90               \\
 5182    & \object{  OC21-M337}&  42 & 177.75772095 & -55.61388397 & 16.698$\pm$  0.003 & 17.610$\pm$  0.002 & 16.643 & 17.538 &  -8.92$\pm$   0.46 & 0 & 5802$\pm$ 243 & $<$   28   &$<$ 2.10 &$<$     2.10               \\
 4367    & \object{  OC21-M281}&  43 & 177.76686096 & -55.63492203 & 17.670$\pm$  0.004 & 18.617$\pm$  0.005 & 17.896 & 18.915 & 167.15$\pm$   1.77 & 0 & 5669$\pm$ 249 & $<$   28   &$<$ 1.64 &$<$     1.71               \\
 5195    & \object{  OC21-M338}&  44 & 177.74420166 & -55.60814285 & 18.028$\pm$  0.014 & 19.097$\pm$  0.011 & 17.973 & 19.025 &  12.22$\pm$   2.51 & 0 & 5241$\pm$ 251 & $<$   47   &$<$ 1.76 &$<$     1.85               \\
 2647    & \object{  OC21-M217}&  45 & 177.79577637 & -55.69689941 & 17.625$\pm$  0.004 & 18.434$\pm$  0.003 & 17.945 & 18.857 &  17.85$\pm$   2.41 & 0 & 6218$\pm$ 243 & $<$   47   &$<$ 2.27 &$<$     2.27               \\
 2250    & \object{   OC21-M35}&  46 & 177.68827820 & -55.69499588 & 17.286$\pm$  0.030 & 18.129$\pm$  0.010 & 17.306 & 18.155 & -18.22$\pm$   2.96 & 1 & 6076$\pm$ 241 &  98$\pm$ 6 & 2.85 &     2.79$\pm$     0.23 \\
 3496    & \object{  OC21-M247}&  47 & 177.81762695 & -55.68233490 & 17.522$\pm$  0.005 & 18.404$\pm$  0.006 & 17.470 & 18.335 &  22.04$\pm$   0.66 & 0 & 5919$\pm$ 241 &  37$\pm$ 4 & 2.21 &     2.20$\pm$     0.25 \\
 3563    & \object{  OC21-M249}&  48 & 177.82002258 & -55.66353226 & 16.936$\pm$  0.004 & 17.954$\pm$  0.003 & 16.884 & 17.885 & -16.30$\pm$   0.80 & 1 & 5414$\pm$ 248 & $<$   28   &$<$ 1.72 &$<$     1.78               \\
 3391    & \object{  OC21-M238}&  49 & 177.78805542 & -55.67609024 & 15.902$\pm$  0.002 & 16.687$\pm$  0.002 & 16.058 & 16.894 &  -6.97$\pm$   0.97 & 0 & 6321$\pm$ 240 &  53$\pm$ 2 & 2.51 &     2.48$\pm$     0.21 \\
 3465    & \object{  OC21-M241}&  51 & 177.78099060 & -55.66153717 & 16.630$\pm$  0.002 & 17.403$\pm$  0.002 & 16.786 & 17.610 &  -1.05$\pm$   0.61 & 0 & 6373$\pm$ 240 & $<$   32   &$<$ 2.41 &$<$     2.38               \\
 3560    & \object{  OC21-M248}&  52 & 177.79272461 & -55.66469574 & 17.322$\pm$  0.006 & 18.138$\pm$  0.003 & 17.270 & 18.069 & -23.36$\pm$   1.15 & 1 & 6189$\pm$ 239 & $<$    9   &$<$ 1.94 &$<$     1.91               \\
 4499    & \object{  OC21-M285}&  53 & 177.80053711 & -55.65328217 & 17.676$\pm$  0.006 & 18.535$\pm$  0.005 & 17.909 & 18.843 & -20.39$\pm$   1.36 & 1 & 6011$\pm$ 244 &  57$\pm$15 & 2.19 &     2.21$\pm$     0.45 \\
 1640    & \object{  OC21-M184}&  54 & 177.77345276 & -55.72911453 & 16.947$\pm$  0.003 & 17.750$\pm$  0.003 & 17.026 & 17.855 &   0.26$\pm$   0.88 & 0 & 6243$\pm$ 240 &  30$\pm$ 5 & 2.23 &     2.21$\pm$     0.31 \\
 1582    & \object{  OC21-M181}&  55 & 177.78698730 & -55.72935104 & 17.707$\pm$  0.003 & 18.683$\pm$  0.009 & 17.786 & 18.788 &  -5.78$\pm$   0.92 & 0 & 5563$\pm$ 248 &  21$\pm$ 4 & 1.23 &     1.30$\pm$     0.66 \\
 1605    & \object{  OC21-M182}&  56 & 177.77871704 & -55.72304916 & 16.432$\pm$  0.002 & 17.309$\pm$  0.013 & 16.511 & 17.414 & -15.61$\pm$   0.66 & 1 & 5939$\pm$ 243 &  12$\pm$ 2 & 1.13 &     1.15$\pm$      -- \\
 1726    & \object{  OC21-M185}&  57 & 177.79896545 & -55.72079086 & 17.340$\pm$  0.004 & 18.186$\pm$  0.004 & 17.303 & 18.137 &  55.35$\pm$  14.08 & 0 & 6064$\pm$ 240 & $<$   32   &$<$ 2.37 &$<$     2.34               \\
 2436    & \object{  OC21-M202}&  58 & 177.78025818 & -55.71140289 & 17.424$\pm$  0.007 & 18.315$\pm$  0.004 & 17.786 & 18.794 & -26.33$\pm$   1.01 & 1 & 5884$\pm$ 248 &  19$\pm$ 1 & 1.10 &     1.17$\pm$     0.35 \\
 2466    & \object{  OC21-M205}&  59 & 177.78466797 & -55.70513535 & 17.080$\pm$  0.002 & 17.840$\pm$  0.003 & 17.442 & 18.319 & -22.22$\pm$   1.26 & 1 & 6429$\pm$ 242 &  98$\pm$ 9 & 2.75 &     2.70$\pm$     0.26 \\
 2599    & \object{  OC21-M213}&  60 & 177.79322815 & -55.70159912 & 17.025$\pm$  0.003 & 17.811$\pm$  0.003 & 17.345 & 18.234 & -21.72$\pm$   4.03 & 1 & 6316$\pm$ 242 &  41$\pm$ 6 & 2.17 &     2.17$\pm$     0.29 \\
 3494    & \object{  OC21-M246}&  61 & 177.79835510 & -55.68308640 & 16.093$\pm$  0.004 & 16.920$\pm$  0.004 & 16.041 & 16.851 &   6.53$\pm$   0.82 & 0 & 6143$\pm$ 239 &  60$\pm$ 2 & 2.69 &     2.64$\pm$     0.20 \\
 3481    & \object{  OC21-M245}&  62 & 177.78889465 & -55.68545532 & 16.942$\pm$  0.004 & 17.755$\pm$  0.004 & 17.098 & 17.962 & -55.02$\pm$   1.12 & 0 & 6201$\pm$ 241 &  95$\pm$ 5 & 2.78 &     2.73$\pm$     0.23 \\
 2349    & \object{   OC21-M48}&  64 & 177.70640564 & -55.70300293 & 17.812$\pm$  0.011 & 18.793$\pm$  0.006 & 17.952 & 18.978 &  24.90$\pm$   2.28 & 0 & 5545$\pm$ 249 & $<$   11   &$<$ 1.18 &$<$     1.26               \\
  620    & \object{  OC21-M137}&  65 & 177.68751526 & -55.74435806 & 17.239$\pm$  0.004 & 18.121$\pm$  0.019 & 17.251 & 18.137 &  -0.72$\pm$   0.80 & 0 & 5919$\pm$ 242 & $<$   19   &$<$ 1.95 &$<$     1.95               \\
 2189    & \object{   OC21-M26}&  67 & 177.67781067 & -55.71120071 & 16.037$\pm$  0.015 & 16.723$\pm$  0.011 & 16.057 & 16.749 & -13.66$\pm$  20.71 & 1 & 6765$\pm$ 235 & $<$   10   &$<$ 2.13 &$<$     2.08               \\
  867    & \object{  OC21-M144}&  68 & 177.75404358 & -55.74374390 & 17.154$\pm$  0.007 & 17.933$\pm$  0.015 & 17.142 & 17.917 & -18.09$\pm$   1.33 & 1 & 6346$\pm$ 238 &  55$\pm$ 1 & 2.76 &     2.70$\pm$     0.15 \\
  855    & \object{  OC21-M143}&  69 & 177.74540710 & -55.74771881 & 17.948$\pm$  0.004 & 18.879$\pm$  0.017 & 17.936 & 18.863 & -27.03$\pm$   0.78 & 0 & 5729$\pm$ 244 & $<$   11   &$<$ 1.57 &$<$     1.60               \\
 1541    & \object{  OC21-M177}&  70 & 177.76091003 & -55.74035263 & 17.350$\pm$  0.006 & 18.202$\pm$  0.010 & 17.429 & 18.307 &  -7.65$\pm$  10.59 & 0 & 6040$\pm$ 242 & $<$   12   &$<$ 1.78 &$<$     1.78               \\
 2282    & \object{   OC21-M41}&  71 & 177.67535400 & -55.69907761 & 16.294$\pm$  0.002 & 17.015$\pm$  0.002 & 16.314 & 17.041 & -11.62$\pm$   3.35 & 0 & 6604$\pm$ 236 & $<$   10   &$<$ 2.16 &$<$     2.11               \\
 1553    & \object{  OC21-M178}&  72 & 177.77110291 & -55.73638916 & 17.212$\pm$  0.008 & 18.005$\pm$  0.007 & 17.291 & 18.110 & -22.26$\pm$   1.08 & 1 & 6286$\pm$ 239 &  89$\pm$ 2 & 2.89 &     2.82$\pm$     0.20 \\
  843    & \object{  OC21-M401}&  73 & 177.75950623 & -55.75041580 & 16.973$\pm$  0.002 & 17.806$\pm$  0.008 & 16.961 & 17.790 &  -7.78$\pm$   0.79 & 0 & 6118$\pm$ 240 & $<$   10   &$<$ 1.86 &$<$     1.84               \\
 1452    & \object{  OC21-M169}&  75 & 177.69763184 & -55.73071289 & 17.872$\pm$  0.005 & 18.852$\pm$  0.010 & 17.842 & 18.813 &   3.18$\pm$   0.99 & 0 & 5548$\pm$ 246 & $<$   10   &$<$ 1.38 &$<$     1.43               \\
  599    & \object{  OC21-M134}&  76 & 177.65110779 & -55.75020599 & 17.454$\pm$  0.004 & 18.393$\pm$  0.004 & 17.466 & 18.409 &  22.26$\pm$   0.98 & 0 & 5699$\pm$ 245 & $<$   10   &$<$ 1.47 &$<$     1.51               \\
  613    & \object{  OC21-M136}&  77 & 177.64176941 & -55.74531937 & 16.530$\pm$  0.002 & 17.242$\pm$  0.003 & 16.542 & 17.258 & -20.79$\pm$   2.03 & 2 & 6645$\pm$ 236 &            &         &                           \\
  141    & \object{  OC21-M353}&  78 & 177.67767334 & -55.77682877 & 17.291$\pm$  0.027 & 18.179$\pm$  0.006 & 17.012 & 17.810 &  -2.31$\pm$   2.87 & 0 & 5895$\pm$ 239 & $<$   13   &$<$ 2.10 &$<$     2.07               \\
  198    & \object{  OC21-M359}&  79 & 177.69346619 & -55.78539276 & 18.068$\pm$  0.006 & 19.065$\pm$  0.012 & 17.924 & 18.875 & -20.60$\pm$   0.98 & 1 & 5487$\pm$ 245 &  65$\pm$ 9 & 2.21 &     2.23$\pm$     0.30 \\
  236    & \object{  OC21-M362}&  81 & 177.71144104 & -55.77384567 & 17.295$\pm$  0.003 & 18.142$\pm$  0.003 & 17.151 & 17.952 & -23.75$\pm$   1.04 & 1 & 6060$\pm$ 239 &  65$\pm$ 1 & 2.76 &     2.70$\pm$     0.20 \\
 2415    & \object{   OC21-M58}&  82 & 177.69410706 & -55.70250320 & 16.188$\pm$  0.005 & 16.935$\pm$  0.005 & 16.328 & 17.120 & -17.04$\pm$   2.88 & 1 & 6487$\pm$ 238 & $<$    5   &$<$ 1.66 &$<$     1.63               \\
 2362    & \object{   OC21-M52}&  83 & 177.71514893 & -55.70048523 & 17.259$\pm$  0.009 & 18.252$\pm$  0.004 & 17.399 & 18.437 & -35.63$\pm$   4.61 & 0 & 5502$\pm$ 250 &  51$\pm$ 7 & 1.72 &     1.80$\pm$     0.33 \\
 2198    & \object{   OC21-M29}&  84 & 177.67022705 & -55.70906830 & 17.090$\pm$  0.007 & 17.921$\pm$  0.005 & 17.110 & 17.947 & -20.57$\pm$   1.19 & 1 & 6126$\pm$ 240 &  88$\pm$ 4 & 2.82 &     2.76$\pm$     0.22 \\
 2082    & \object{  OC21-M200}&  86 & 177.60665894 & -55.71215439 & 17.730$\pm$  0.005 & 18.640$\pm$  0.010 & 17.621 & 18.496 &  16.29$\pm$   2.03 & 0 & 5810$\pm$ 242 & $<$   14   &$<$ 1.84 &$<$     1.84               \\
  352    & \object{  OC21-M375}&  87 & 177.57582092 & -55.76445770 & 17.765$\pm$  0.004 & 18.815$\pm$  0.007 & 17.482 & 18.441 & -16.76$\pm$   0.70 & 1 & 5305$\pm$ 246 & $<$    9   &$<$ 1.36 &$<$     1.41               \\
  608    & \object{  OC21-M135}&  88 & 177.67390442 & -55.74694443 & 16.737$\pm$  0.005 & 17.586$\pm$  0.002 & 16.749 & 17.602 & -21.75$\pm$   0.76 & 1 & 6052$\pm$ 241 &  83$\pm$ 5 & 2.73 &     2.68$\pm$     0.23 \\
 2077    & \object{   OC21-M14}&  89 & 177.62500000 & -55.71352005 & 17.924$\pm$  0.005 & 19.005$\pm$  0.006 & 17.815 & 18.861 &  11.37$\pm$   0.68 & 0 & 5202$\pm$ 250 & $<$   54   &$<$ 1.85 &$<$     1.93               \\
 2112    & \object{   OC21-M16}&  91 & 177.63342285 & -55.70466995 & 17.508$\pm$  0.004 & 18.450$\pm$  0.004 & 17.399 & 18.306 & -20.72$\pm$   3.14 & 1 & 5688$\pm$ 243 &  66$\pm$ 1 & 2.38 &     2.38$\pm$     0.23 \\
 2242    & \object{   OC21-M34}&  93 & 177.66357422 & -55.69713211 & 17.902$\pm$  0.005 & 18.892$\pm$  0.006 & 17.922 & 18.918 &  12.63$\pm$   1.32 & 0 & 5512$\pm$ 248 &  49$\pm$ 9 & 1.87 &     1.93$\pm$     0.37 \\
 1131    & \object{  OC21-M150}&  95 & 177.58456421 & -55.72017670 & 17.111$\pm$  0.003 & 18.131$\pm$  0.003 & 16.831 & 17.760 & -27.53$\pm$   0.92 & 0 & 5407$\pm$ 244 & $<$   28   &$<$ 1.97 &$<$     1.99               \\
 1007    & \object{  OC21-M416}&  96 & 177.50790405 & -55.72454453 & 17.978$\pm$  0.004 & 19.066$\pm$  0.006 & 16.946 & 17.702 & 144.01$\pm$   1.26 & 0 & 5180$\pm$ 237 & $<$   18   &$<$ 2.38 &$<$     2.33               \\
 1307    & \object{  OC21-M160}&  97 & 177.64410400 & -55.73705292 & 17.461$\pm$  0.007 & 18.390$\pm$  0.006 & 17.483 & 18.420 & -23.74$\pm$   1.87 & 1 & 5737$\pm$ 245 &  52$\pm$ 5 & 2.13 &     2.15$\pm$     0.27 \\
  344    & \object{  OC21-M374}& 100 & 177.54367065 & -55.76548004 & 17.152$\pm$  0.018 & 17.810$\pm$  0.119 & 16.869 & 17.436 &  -4.26$\pm$   1.24 & 0 & 6897$\pm$ 231 &   8$\pm$ 0 & 1.87 &     1.80$\pm$     0.10 \\
  630    & \object{  OC21-M138}& 101 & 177.68962097 & -55.75462341 & 17.671$\pm$  0.005 & 18.581$\pm$  0.007 & 17.683 & 18.597 &   9.69$\pm$   1.74 & 0 & 5810$\pm$ 244 & $<$   30   &$<$ 2.06 &$<$     2.07               \\
 2141    & \object{   OC21-M19}& 103 & 177.59860229 & -55.69361877 & 18.020$\pm$  0.007 & 18.961$\pm$  0.006 & 17.911 & 18.817 &  -5.53$\pm$   1.21 & 0 & 5692$\pm$ 243 & $<$   13   &$<$ 1.71 &$<$     1.73               \\
 1328    & \object{  OC21-M162}& 106 & 177.64506531 & -55.73122406 & 16.789$\pm$  0.002 & 17.697$\pm$  0.003 & 16.811 & 17.727 &  17.82$\pm$   0.34 & 0 & 5817$\pm$ 244 & $<$   13   &$<$ 1.67 &$<$     1.69               \\
 2048    & \object{  OC21-M196}& 107 & 177.54794312 & -55.69825745 & 17.456$\pm$  0.004 & 18.341$\pm$  0.008 & 17.059 & 17.816 & -69.91$\pm$   1.16 & 0 & 5907$\pm$ 237 &  49$\pm$ 3 & 2.75 &     2.69$\pm$     0.15 \\
 2039    & \object{  OC21-M195}& 108 & 177.55949402 & -55.69301224 & 17.194$\pm$  0.018 & 18.117$\pm$  0.005 & 16.797 & 17.592 &   0.51$\pm$   1.21 & 0 & 5760$\pm$ 238 &  35$\pm$ 2 & 2.45 &     2.41$\pm$     0.21 \\
 1996    & \object{  OC21-M192}& 109 & 177.55455017 & -55.70639038 & 17.085$\pm$  0.003 & 17.939$\pm$  0.009 & 16.688 & 17.414 & -20.18$\pm$   1.21 & 1 & 6032$\pm$ 236 & $<$   10   &$<$ 2.14 &$<$     2.09               \\
 1029    & \object{  OC21-M419}& 110 & 177.48760986 & -55.71768188 & 18.232$\pm$  0.005 & 19.350$\pm$  0.010 & 17.200 & 17.986 &   4.69$\pm$   1.11 & 0 & 5085$\pm$ 238 & $<$   16   &$<$ 2.22 &$<$     2.19               \\
 1991    & \object{  OC21-M190}& 111 & 177.57115173 & -55.70736694 & 18.154$\pm$  0.010 & 19.256$\pm$  0.007 & 17.757 & 18.731 &  30.50$\pm$   0.53 & 0 & 5135$\pm$ 246 & $<$   48   &$<$ 2.05 &$<$     2.09               \\
 1834    & \object{  OC21-M443}& 112 & 177.47384644 & -55.70010757 & 18.047$\pm$  0.005 & 19.179$\pm$  0.005 & 17.682 & 18.697 &   3.41$\pm$   0.88 & 0 & 5042$\pm$ 249 & $<$   44   &$<$ 1.86 &$<$     1.93               \\
 2131    & \object{   OC21-M18}& 113 & 177.63011169 & -55.69727707 & 17.749$\pm$  0.007 & 18.706$\pm$  0.007 & 17.640 & 18.562 &  54.13$\pm$   4.03 & 0 & 5632$\pm$ 244 & $<$   23   &$<$ 1.90 &$<$     1.92               \\
 1973    & \object{  OC21-M187}& 114 & 177.56176758 & -55.71268082 & 17.025$\pm$  0.002 & 17.860$\pm$  0.003 & 16.628 & 17.335 & -10.28$\pm$   1.21 & 0 & 6109$\pm$ 235 &  45$\pm$ 4 & 2.77 &     2.70$\pm$     0.11 \\
 2747    & \object{  OC21-M471}& 116 & 177.47677612 & -55.66456223 & 17.266$\pm$  0.003 & 18.233$\pm$  0.004 & 17.135 & 18.059 & -13.38$\pm$   1.38 & 1 & 5595$\pm$ 244 &  27$\pm$ 3 & 1.78 &     1.80$\pm$     0.30 \\
 2909    & \object{  OC21-M224}& 117 & 177.57012939 & -55.67611313 & 17.402$\pm$  0.003 & 18.356$\pm$  0.004 & 17.543 & 18.543 &  -5.23$\pm$   0.67 & 0 & 5643$\pm$ 248 &  45$\pm$ 3 & 1.81 &     1.87$\pm$     0.27 \\
 3803    & \object{  OC21-M508}& 118 & 177.48970032 & -55.65198517 & 16.499$\pm$  0.008 & 17.235$\pm$  0.004 & 16.387 & 17.087 & -17.68$\pm$   1.61 & 1 & 6536$\pm$ 235 &  46$\pm$ 7 & 2.79 &     2.72$\pm$     0.18 \\
 2724    & \object{  OC21-M467}& 119 & 177.47935486 & -55.68232727 & 18.121$\pm$  0.008 & 19.156$\pm$  0.013 & 17.990 & 18.982 &  20.55$\pm$   2.15 & 0 & 5355$\pm$ 247 & $<$   38   &$<$ 1.89 &$<$     1.94               \\
 2953    & \object{  OC21-M230}& 120 & 177.56149292 & -55.66583252 & 16.318$\pm$  0.002 & 17.011$\pm$  0.003 & 16.459 & 17.198 & -19.10$\pm$   1.80 & 1 & 6733$\pm$ 236 &  44$\pm$ 7 & 2.75 &     2.68$\pm$     0.20 \\
 2171    & \object{   OC21-M22}& 121 & 177.59326172 & -55.68820572 & 18.074$\pm$  0.007 & 18.989$\pm$  0.010 & 17.965 & 18.845 & -10.22$\pm$   1.34 & 0 & 5790$\pm$ 242 &  84$\pm$ 5 & 2.64 &     2.61$\pm$     0.24 \\
 3822    & \object{  OC21-M253}& 122 & 177.55290222 & -55.65555191 & 17.481$\pm$  0.006 & 18.448$\pm$  0.005 & 17.627 & 18.642 &   1.75$\pm$   0.53 & 0 & 5595$\pm$ 249 & $<$   22   &$<$ 1.56 &$<$     1.63               \\
 3112    & \object{   OC21-M74}& 123 & 177.59184265 & -55.68222809 & 17.257$\pm$  0.006 & 18.203$\pm$  0.007 & 17.166 & 18.083 &  17.36$\pm$   0.85 & 0 & 5673$\pm$ 244 & $<$   26   &$<$ 1.98 &$<$     2.00               \\
 2883    & \object{  OC21-M219}& 124 & 177.56431580 & -55.68196869 & 17.644$\pm$  0.005 & 18.675$\pm$  0.007 & 17.785 & 18.862 &   2.83$\pm$   4.57 & 0 & 5369$\pm$ 252 & $<$   38   &$<$ 1.57 &$<$     1.68               \\
 4775    & \object{  OC21-M548}& 126 & 177.57238770 & -55.60731888 & 16.782$\pm$  0.002 & 17.703$\pm$  0.004 & 16.642 & 17.518 &  40.61$\pm$   0.55 & 0 & 5767$\pm$ 242 & $<$   29   &$<$ 2.18 &$<$     2.17               \\
 3881    & \object{  OC21-M264}& 127 & 177.57377625 & -55.64166260 & 16.894$\pm$  0.005 & 17.722$\pm$  0.006 & 17.040 & 17.916 &  36.76$\pm$   0.60 & 0 & 6138$\pm$ 242 &  42$\pm$ 9 & 2.24 &     2.23$\pm$     0.38 \\
 4638    & \object{  OC21-M535}& 128 & 177.53495789 & -55.61775970 & 17.848$\pm$  0.004 & 18.738$\pm$  0.005 & 17.823 & 18.705 & -23.78$\pm$   5.09 & 1 & 5888$\pm$ 242 &  58$\pm$ 8 & 2.39 &     2.38$\pm$     0.28 \\
 3094    & \object{   OC21-M70}& 129 & 177.61021423 & -55.66329956 & 17.292$\pm$  0.003 & 18.192$\pm$  0.003 & 17.201 & 18.072 &  -6.54$\pm$   0.88 & 0 & 5848$\pm$ 242 &  19$\pm$ 5 & 1.74 &     1.74$\pm$     0.71 \\
 4679    & \object{  OC21-M541}& 130 & 177.50422668 & -55.61781311 & 16.557$\pm$  0.026 & 17.255$\pm$  0.003 & 16.532 & 17.222 & -15.43$\pm$   2.70 & 1 & 6709$\pm$ 235 &  38$\pm$ 5 & 2.69 &     2.62$\pm$     0.15 \\
 4595    & \object{  OC21-M530}& 131 & 177.52270508 & -55.62817383 & 16.615$\pm$  0.003 & 17.361$\pm$  0.003 & 16.590 & 17.328 &  16.29$\pm$   3.27 & 0 & 6491$\pm$ 236 &  47$\pm$ 2 & 2.79 &     2.72$\pm$     0.10 \\
 3767    & \object{  OC21-M504}& 133 & 177.50112915 & -55.63995743 & 17.890$\pm$  0.007 & 18.876$\pm$  0.007 & 17.778 & 18.728 &  -8.20$\pm$   0.76 & 0 & 5527$\pm$ 245 & $<$   40   &$<$ 2.05 &$<$     2.08               \\
 4764    & \object{  OC21-M547}& 134 & 177.53811646 & -55.61055374 & 17.980$\pm$  0.005 & 18.811$\pm$  0.005 & 17.840 & 18.626 &  13.99$\pm$   2.50 & 0 & 6126$\pm$ 238 &  47$\pm$ 5 & 2.63 &     2.58$\pm$     0.22 \\
 3668    & \object{  OC21-M491}& 135 & 177.48237610 & -55.65921021 & 18.026$\pm$  0.021 & 19.068$\pm$  0.007 & 17.707 & 18.647 & -17.28$\pm$   0.89 & 1 & 5331$\pm$ 245 & $<$   16   &$<$ 1.66 &$<$     1.69               \\
\end{longtable}
\end{landscape}
}